\documentclass[12pt]{iopart}
\usepackage{iopams}  
\usepackage{graphicx}
\usepackage{subfigure}

\usepackage[dvipsnames]{xcolor}

\begin{document}

\title[Uniformity Transition]{Uniformity Transition for Ray Intensities in Random Media}

\author{Marc Pradas$^{1}$, Alain Pumir$^{2}$ and  Michael Wilkinson$^{1}$}

\address{$^1$ School of Mathematics and Statistics,
The Open University, Walton Hall, Milton Keynes, MK7 6AA, England,\\
$^2$ Laboratoire de Physique,
 Ecole Normale Sup\'erieure de Lyon, CNRS, Universit\'e de Lyon,
 F-69007, Lyon, France,}
\ead{marc.pradas@open.ac.uk,
alain.pumir@ens-lyon.fr,
m.wilkinson@open.ac.uk}
\vspace{10pt}
\begin{indented}
\item November 2017
\end{indented}

\begin{abstract}
This paper analyses a model for the intensity of distribution for rays 
propagating without absorption in a random medium. The random medium 
is modelled as a dynamical map. After $N$ iterations, the intensity is modelled as 
a sum $S$ of ${\cal N}$ contributions from different trajectories, each of which is 
a product of $N$ independent identically distributed random variables $x_k$, representing
successive focussing or de-focussing events. The number of ray 
trajectories reaching a given point is assumed to proliferate
exponentially: ${\cal N}=\Lambda^N$, for some $\Lambda>1$. 
We investigate the probability distribution of $S$. We find a phase transition as 
parameters of the model are varied. There is a phase where the 
fluctuations of $S$ are suppressed as $N\to \infty$, and a phase where 
the $S$ has large fluctuations, for which we provide a large deviation 
analysis.
\end{abstract}

%\submitto{\jpa}
\maketitle

\section{Introduction}
\label{sec: 1}

We consider a model for light rays propagating through a random medium with
negligible absorption.  Random fluctuations of the refractive index cause rays to 
diverge or to focus, leading to fluctuations of the light intensity, depending 
on the path of the light ray to reach the point where the intensity is 
observed \cite{Tat61,Jak+88,Usc68}. Because the effects of each 
successive focussing or de-focussing events are to multiply the light intensity 
by a random factor, the effects of focussing are expected to increase exponentially 
with the path length. On the other hand, the intensity at a given point is the sum of the 
intensities from all of the rays reaching that point. The number of rays reaching a 
point increases exponentially with the path length, and we expect that the proliferation
of rays will tend to average out the fluctuations of the intensity.  There are, therefore, 
two effects on the distribution of intensity fluctuations which compete as we increase  
the path length. Does the effect of focussing along individual rays dominate, so that 
the light intensity shows an increasingly pronounced speckle? Or does the proliferation 
of paths become dominant, so that intensity fluctuations are averaged out and the 
medium behaves as a diffuser which produces a uniform intensity? In this paper we
introduce and analyse a very simplified, physically well-motivated model, which is analytically solvable.
We show that this model has a phase transition between a fluctuation-dominated phase 
and a uniform phase.  

\section{A model for intensity statistics}
\label{sec: 2}

Several approaches have been proposed to compute the distribution of 
the intensity of waves travelling through a random medium. 
Many authors have treated the solution of the wave equation directly, 
see e.g.~\cite{Sha86,Mar+88}. Others have 
simplified the problem by considering a short-wavelength limit and 
concentrating on the ray trajectories \cite{Usc68,Han82,Met+14}. This approach 
relates the high-intensity events to the effects of focussing, and makes elegant connections with 
catastrophe theory \cite{Ber77,Ber81}. The use of catastrophe theory is appropriate when only a 
few rays reach each observation point. As we move deeper into a random medium, 
however, the number of trajectories which can reach a given point proliferates, 
essentially exponentially. It is this case which is addressed in our work: we consider 
propagation with negligible absorption in a short-wavelength limit, so that the intensities 
are determined by focussing of rays, but the number of rays which could contribute 
is extremely large. Our objective in this paper is to analyse a 
solvable model which can serve as a benchmark for future studies of 
more specific models.
 
We motivate our model by considering a simplified 
one-dimensional problem of ray propagation along the $z$ axis. The point at which a 
ray crosses the perpendicular axis after propagation for a distance of $z=n\Delta z$
(where $\Delta z$ is some fixed increment) is $x_n$. The evolution of the ray 
position $x_n$ is described by a sequence of random one-dimensional maps, $f_n$:
\begin{equation}
x_{n+1} = f_n(x_n)
\ .
\end{equation}
We assume that this random dynamical system has 
\lq chaotic' properties with a positive Lyapunov exponent
\cite{Ott02}. The density of initial conditions is $\rho_0$, and 
the density of trajectories after $N$ iterations of the map is denoted as $\rho_N(x)$. 
If the map were invertible, the density would be $\rho_0(x_N)/F'_N(x_N)$, where 
$F_N(x)$ is the mapping for $N$ iterations so that
\begin{equation}
\label{eq: 2.1}
F'_N(x)=\left(\frac{\partial x_N}{\partial x_0}\right)
\end{equation}
is the stability factor of the trajectory, $x_N(x)$ is the $N$ step pre-image of $x$, and 
$\rho_0(x)$ is the initial density at $x$. Usually, however, a point will have multiple pre-images, so that
\begin{equation}
\label{eq: 2.2}
\rho_N(x)=\sum_{j=1}^{\mathcal{N}} \frac{\rho_0(x_j)}{|F_N'(x_j)|}
\end{equation}
where the $x_j$ are the ${\cal N}$ pre-images of $x$. The number of pre-images 
of a point is expected to proliferate exponentially (with exponent equal to the 
topological entropy \cite{Nit+99}), and after $N$ iterations we have:
\begin{equation}
\label{eq: 2.3}
{\cal N}\sim \Lambda ^N
\end{equation}
for some constant $\Lambda>1$. The stability factor of the 
trajectory is a product of terms for each time step, 
where the sum runs over all of the pre-images of $x$ at $n=0$ and 
the sensitivity of each trajectory is a product of independent
terms: 
\begin{equation}
\label{eq: 2.4}
F_N'(x)=\prod_{k=1}^N \left| \frac{\partial x_{k}}{\partial x_{k-1}} \right|_{x_{k-1}}
=\prod_{k=1}^N f'_k(x_k)
\end{equation}
where the $x_k$ are the successive pre-images after $k$ iterations. 
When $N$ is large, the density of trajectories is therefore constructed as a 
sum of a large number of terms, each of which is the product of a large number of factors. 

The analysis of how $\rho_N(x)$ varies as a function of $x$ for a specific system 
is clearly a difficult and usually intractable problem. However, the large number of 
proliferating pre-images implies that a statistical approach may yield valuable insights.
In this paper we consider a statistical model for the density, represented by a sum $S$,
which is constructed using a set of independent, identical distributed variables, $x_k$.
The model is defined by the equations
\begin{eqnarray}
\label{eq: 2.5}
{\cal N}&=&{\rm int}(\Lambda^N)
\nonumber \\
S&=&\sum_{j=1}^{\cal N} X_j
\nonumber \\
X_j&=&\prod_{k=1}^N x_k\ .
\end{eqnarray}
Because intensity is a positive quantity, we assume that all of the factors 
$x_k$ are positive.
The problem is to characterise the probability distribution of $S$ in the limit 
$N\gg1$, given the value of $\Lambda$ and the probability density function (PDF) of $x_k$.
If $S$ approaches a limit with small fluctuations relative to its 
magnitude, the density at large times is uniform. Alternatively, if the 
fluctuations of $S$ relative to its size grow, then the density
becomes highly inhomogeneous. 

\begin{figure}[h t b]
\centering
\includegraphics[width=0.99\textwidth]{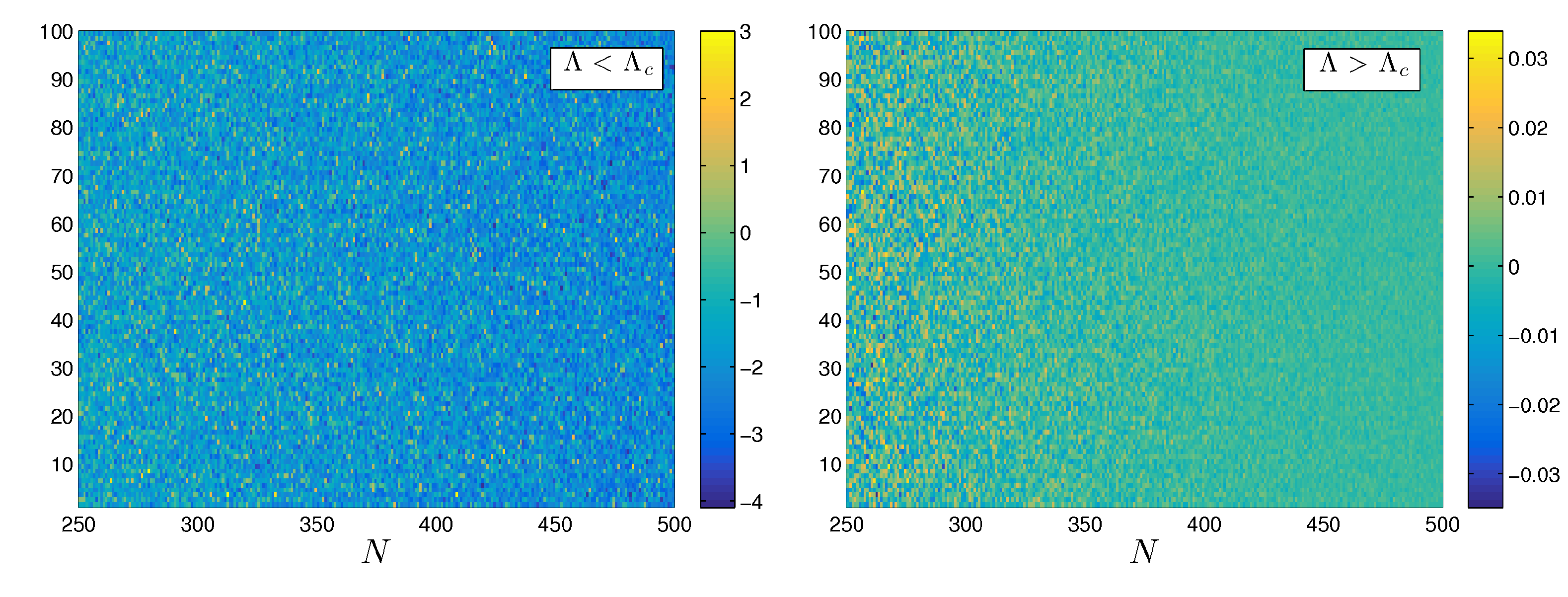}
\caption{Representation of a set of values of the random variable $S$ allocated 
in the vertical axis over different values of $N$ (horizontal axis). Left and right panels 
correspond to a value of the parameter $\Lambda$ below 
($\Lambda = 0.99\Lambda_{\rm c}$) and above ($\Lambda = 1.01\Lambda_{\rm c}$) the critical point 
$\Lambda_{\rm c}$, respectively. The colour coding illustrates from low (blue) 
to high (yellow)  values of $S$, and the colour bar is in decimal logarithmic scale.
At $N = 500$, the largest (smallest) fluctuations of $S$ are larger 
(smaller) than the mean by a factor $10^3$ ($10^{-3}$) 
for the left panel, which shows the extreme nature of the fluctuations.
On the contrary, the fluctuations do not exceed the mean by more 
than $\approx 10\%$ for the right panel.
}
\label{fig: 1}
\end{figure}

We note that in the model described by Eq.~(\ref{eq: 2.5}) there are competing effects. 
The fact that the $X_j$ are a product of many factors implies that they have very 
wide fluctuations in magnitude. On the other hand, $S$ is a sum of an exponentially 
large number of independent quantities, so that fluctuations may be averaged away. 
We must consider which effect dominates, and whether, in the limit as $N\to \infty$, 
the dominant effect can change as the parameter $\Lambda$ is varied. In the following 
we show that there is a phase transition: when $\Lambda$ is relatively small, $S$ shows very large fluctuations, but as $\Lambda $ is increased beyond a critical value 
$\Lambda_{\rm c}$, the fluctuations of $S$ in the limit 
as $N\to \infty$ are suddenly suppressed. 
A numerical illustration of this effect is shown in Fig.~\ref{fig: 1}, where we can see how a set 
$\{S_1,\dots,S_m\}$, with $m=100$, evolves as we increase $N$ for two different 
values of $\Lambda$. When $\Lambda<\Lambda_{\rm c}$ the random variable $S$ 
exhibits inhomogeneous fluctuations spanning several decades in magnitude, 
while these fluctuations are largely suppressed when 
$\Lambda>\Lambda_{\rm c}$. For this numerical example the random 
variables $x_k$ in Eq.~(\ref{eq: 2.5}) are drawn from a log-normal distribution 
(see Section~\ref{sec: 6} for more details).

The model given by Eq.~(\ref{eq: 2.5}) appears to be quite realistic as 
a model for fluctuations of ray intensity: if rays reach the point of 
observation via chaotic trajectories, then it is plausible that these rays 
will sample different regions of the random medium and that the intensity 
factors will be independent. The most significant weakness of our model 
is that it does not represent the effects of propagation: the intensity predicted 
by the model after $N+1$ steps is un-related to the realisation of the 
model for $N$ steps. A more realistic model may take account of the cumulative 
effect of focusing along paths. However, if we are interested in the 
distribution at a single point, there are no obvious reasons 
why the predictions of our model should be suspect. In addition,  
our model has the advantage of being highly amenable to analytical investigations.
   
We note that the model presented here is somewhat analogous to models for the 
partition function of the Ising model and other interacting spin systems 
on disordered Bethe lattices \cite{Bet35,Bax82}. 
We remark, however, that we are aiming at
a different type of result: our quantity $S$ is analogous to the partition 
function, and we are concerned with its probability density. This would be 
analogous to studying the probability distribution of the partition function 
under different realisations of the lattice disorder. The model is also 
somewhat reminiscent of Derrida's random energy model for a spin glass 
\cite{Der81}.
Our model is also quite closely related to a model used in studies of hopping 
conductivity \cite{Ngu+85,Spi+96}. That model differs by having random elements 
with different signs (or, more generally, different complex phases), but it also exhibits a 
phase transition, associated with a transition of the sign of the sum. 

\section{Explicit analytical calculations}
\label{sec: 3}

In the following we simplify the discussion by making a specific choice of the 
PDF of the $x_k$. We give these variables a log-normal distribution by writing
\begin{equation}
\label{eq: 2.6}
x_k=\exp(y_k)
\ ,\ \ \ 
P_y=\frac{1}{\sqrt{2\pi}\sigma}\exp\left[-\frac{(y-\mu)^2}{2\sigma^2}\right]
\end{equation}
where $\mu$ and $\sigma$ are constant 
(throughout, $P_X$ will denote the probability density function of a random 
variable $X$, and $\langle X\rangle$ represents its expectation value).
With this choice of PDF, the moments of $x_k$ 
are obtained explicitly as:
\begin{equation}
\label{eq: 2.7}
\langle x^k \rangle=\exp(k\mu+k^2\sigma^2/2)
\end{equation}
which, as we shall see below, enable us to make explicit calculations. 
Later, we shall show that qualitative results obtained from this distribution are true for a very 
general choice of the probability distribution of the factors $x_k$.

In the following section we describe three calculations that can be done with the 
model (\ref{eq: 2.5}), giving explicit results for the special case where the $x_k$ have a 
log-normal distribution (as defined by (\ref{eq: 2.6})). 

\subsection{Mean value}
\label{sec: 3.1}

The mean value of $S$ is 
\begin{equation}
\label{eq: 3.1.1}
\langle S\rangle\equiv \exp[NZ_0]=\left[\Lambda \langle x\rangle\right]^N
\end{equation}
where the first equality defines the growth exponent $Z_0$. 
Using Eq.~(\ref{eq: 2.7}), we find for the log-normal model
\begin{equation}
\label{eq: 3.1.2}
Z_0=\mu +\ln \Lambda+\frac{1}{2}\sigma^2.
\end{equation}
The parameter $\mu$ can be adjusted to make the mean value of $S$ independent of $N$ 
(which is a physical constraint on the intensity distribution for a non-absorbing medium),
but this is irrelevant to the condition for the phase transition.

\subsection{Normalised central moments}
\label{sec: 3.2}

A central moment of $S$ is $\langle \Delta S^k\rangle$
where $\Delta S=S-\langle S\rangle$. We consider the \emph{normalised 
central moments}
\begin{equation}
\label{eq: 3.2.0}
M_k\equiv \frac{\langle \Delta S^k\rangle}{\langle S\rangle^k}\sim \xi_k^N
\end{equation}
where the second equality defines the growth factor $\xi_k$.
 We find that, in the limit as $N\to\infty$, $M_2\sim \xi_2^N$ with
\begin{equation}
\label{eq: 3.2.2}
\xi_2=\frac{\langle x^2\rangle}{\Lambda\langle x\rangle^2}
\ .
\end{equation}
This implies that the dispersion of the distribution of $S$ approaches zero
as $N\to \infty$ if $\xi_2<1$, suggesting that the distribution will condense 
onto a delta function. This can be generalised. Consider the third moment 
\begin{equation}
\label{eq: 3.2.3}
\langle \Delta S^3\rangle={\cal N}\left[ \langle X^3\rangle-3\langle X^2\rangle\langle X\rangle
-2 \langle X\rangle^3\right]
\ .
\end{equation}
Noting that 
\begin{equation}
\label{eq: 3.2.4}
\langle X^3\rangle=\left[\langle x^3\rangle\right]^N
\ ,\ \ \ 
\langle X^2\rangle\langle X\rangle=\left[\langle x^2\rangle\langle x\rangle\right]^N
\ ,\ \ \ 
\langle X\rangle^3=\left[\langle x\rangle^3\right]^N
\end{equation}
we see that 
\begin{equation}
\label{eq: 3.2.5}
M_3\equiv \frac{\langle\Delta S^3\rangle}{\langle S\rangle^3}\sim \xi_3^N
\end{equation}
where 
\begin{equation}
\label{eq: 3.2.6}
\xi_3=\frac{{\rm max}_{l=0,1,2}\langle x^{3-l}\rangle\langle x\rangle^l}{(\ln \Lambda)^2\langle x\rangle^3}  
\ .
\end{equation}
In general, $\langle \Delta S^k\rangle$ is (for integer $k>1$) a linear combination 
of ${\cal N}$ times $\langle X^{k-l}\rangle\langle X\rangle^l$, with $l=0,\ldots ,k-1$. 
The integer  coefficients are related to Pascal's triangle, but their values are irrelevant
to determining the growth factors $\xi_k$. 
The value of  $\langle \Delta S^k\rangle$ is determined by the largest (in magnitude) 
of the values of $\langle x^{k-l}\rangle \langle x\rangle^l$. We have
\begin{equation}
\label{eq: 3.2.7}
M_k\equiv \frac{\langle \Delta S^k\rangle}{\langle S\rangle^k}\sim \xi_k^N
\end{equation}
where 
\begin{equation}
\label{eq: 3.2.8}
\xi_k=\frac{{\rm max}_{l=0,\ldots k-1}\langle x^{k-l}\rangle\langle x\rangle^l}
{\ln \Lambda^{k-1}\langle x\rangle^k}
\ .
\end{equation}
In general, we cannot conclude that $l=0$ is the largest term, but
for the log-normal model we have an explicit expression (\ref{eq: 2.7}) for 
the expectation values, and we find
\begin{equation}
\label{eq: 3.2.9}
\xi_k=\exp\left[(k-1)\left( k\sigma^2/2-\ln \Lambda \right)\right]
\ .
\end{equation}
This expression has been derived for positive integer values of $k$. It is 
however an analytic function and we can consider the consequences of assuming 
that it is valid for arbitrary values of $k$.

\subsection{Largest element in sum}
\label{sec: 3.3}

We can consider the PDF of $X_{\rm m}$, the largest element of the sum in equation (\ref{eq: 2.5}), 
using a combination of large deviation  \cite{Fre+84,Tou09} and extreme value \cite{Gum35} 
approaches. The distribution of $X$ is more conveniently described in terms of a logarithmic variable
\begin{equation}
\label{eq: 3.3.1}
Y=\frac{1}{N}\ln\,X = \frac{1}{N}\sum_{k=1}^{N} y_k
\end{equation}
where $y_k = \ln\, x_k$. Note that $Y$ is the mean value of $y_k$, 
so that the distribution of $Y$ is expected to be described by a large-deviation ansatz 
\cite{Fre+84,Tou09}:
\begin{equation}
\label{eq: 3.3.2}
P_Y\sim \exp[-NJ(Y)]
\end{equation}
where $J(Y)$ is termed the large deviation \emph{entropy function} 
or \emph{rate function} \cite{Tou09}. 
For the log-normal model the entropy function can be determined explicitly:
\begin{equation}
\label{eq: 3.3.3}
J(Y)=\frac{(Y-\mu)^2}{2\sigma^2}
\ .
\end{equation}
The precise form of the distribution of the maximal value of $Y$, namely 
$Y_{\rm m}=\ln X_{\rm m}/N$  is then 
determined from the Gumbel distribution \cite{Gum35}. However the 
essential features are easily explained. The peak of the distribution of $Y_{\rm m}$ is 
at position $Y_0$, determined by the condition that the product of the probability density 
and the number of samples is of order unity:
\begin{equation}
\label{eq: 3.3.0}
{\cal N}P_Y({Y_0})\sim 1
\ .
\end{equation}
By using ${\cal N} \sim \Lambda^N$, the above condition can be expressed in terms 
of the entropy function as $J(Y_0)=\ln\, \Lambda$, 
which has two solutions. Of these we must consider the larger solution, because we 
are considering the distribution of maximal values. 
For the log-normal model this gives
\begin{equation}
\label{eq: 3.3.4}
Y_0=\mu+\sqrt{2\sigma^2\ln \Lambda}
\ .
\end{equation}
The probability density to obtain $Y_{\rm m}<Y_0$ is extremely small: exponentials 
of exponentials. The probability density for $Y_{\rm m}>Y_0$ is approximately \cite{Gum35}
\begin{equation}
\label{eq: 3.3.5}
P_{Y_{\rm m}}(Y_{\rm m})\sim {\cal N}\exp[-NJ(Y_{\rm m})]
\ .
\end{equation}
When $Y_{\rm m}-Z_0$ is sufficiently small, we can approximate this using a Taylor 
expansion about $Z_0$. The derivative of $J(Y)$ at $Z_0$ is 
\begin{equation}
\label{eq: 3.3.6}
J'(Z_0)=\frac{Z_0-\mu}{\sigma^2}=\frac{1}{2}+\frac{\ln \Lambda}{\sigma^2}
\end{equation}
so that $P_{Y_{\rm m}}\sim \exp[-\alpha N(Y_{\rm m}-Z_0)]$ with
\begin{equation}
\label{eq: 3.3.7}
\alpha=\frac{1}{2}+\frac{\ln \Lambda}{\sigma^2}
\ .
\end{equation}
Therefore, the corresponding PDF of $X_{\rm m}$ is
\begin{equation}
\label{eq: 3.3.8}
P_{X_{\rm m}}\sim X_{\rm m}^{-(1+\alpha)}
\ .
\end{equation}
We remark that the case where $\alpha=1$ may be significant. If $\alpha<1$, 
the approximation (\ref{eq: 3.3.8}) suggests that the integral determining the mean value 
is divergent. In this case the mean value is determined by the behaviour of the 
tail of the distribution at values much larger than the typical value of $X$. For our log-normal 
model, the critical point where $\alpha=1$ is determined by the condition 
\begin{equation}
\label{eq: 3.3.9}
\ln \Lambda_{\rm c}=\sigma^2/2
\ .
\end{equation}

\section{Inferences from calculations}
\label{sec: 4}

The calculations discussed in section \ref{sec: 3} can be used to 
infer properties of the distribution $P_S$, as follows.

\subsection{Existence of delta-function measures}
\label{sec: 4.1}

Let us consider the consequences of finding that $\xi_k<1$ for some value of $k$. 
This condition may be satisfied if the distribution $P_S$ approaches a delta function as 
$N\to \infty$. This is also consistent with the distribution $P_S$ having a long \lq tail', 
provided this tail decreases sufficiently rapidly as $S\to \infty$ and as $N\to \infty$. 
For example, a distribution of the form
\begin{equation}
\label{eq: 4.1.1}
P_S\sim [1-w(N)]\delta (S-S_0)+w(N)\Theta(S-S_0)(S-S_0)^{-(1+\alpha)},
\end{equation}
with $w(N)\sim \exp[-\beta N]$, $\alpha>k$, and $\beta>0$,  is consistent 
with having normalised central moments that go to zero, $M_k\to 0$, as $N\to \infty$. In this sense, 
showing that $\xi_k<1$ for $k>1$ implies that there is a delta-function component of $P_S$ 
that emerges as $N\to \infty$. It is hence desirable to determine the region of parameter space 
for which the delta-function component of $P_S$ is present. 

We have seen that $M_k\sim \xi_k^N$, where, in the log-normal case,
$\xi_k$ is given by expression (\ref{eq: 3.2.9}), 
which is an analytical function of $k$. In the following, we  assume 
that this expression is valid for any real positive value of $k$. 
This is very similar in spirit to the \lq replica trick' 
where the free energy is obtained from the $n^{\rm th}$ moment of the 
partition function by taking the limit as $n\to 0$
\cite{Edw+75,Par84}. Let us determine for which combination of the 
model parameters ($\Lambda$, $\mu$ and $\sigma$) the value of $\xi_k$ may
be less 
than unity for some choice of $k>1$. Clearly the value of $\mu$ is irrelevant because it 
does not appear in (\ref{eq: 3.2.9}). If $\sigma^2/2<\ln\Lambda$, 
then, for all values of $k$, satisfying 
$1 < k < 2 \ln \Lambda/\sigma^2 $, $\xi_k<1$.
This therefore 
suggests that there is a delta-function component whenever 
$\ln \Lambda>\sigma^2/2$.

In the case where $\sigma^2/2>\ln \Lambda$, the values of $\xi_k$ 
can be less than $1$ only for $k < 0$. 
In this case, we cannot infer the existence of a delta-function component.
We have seen that when $\sigma^2/2>\ln \Lambda$, the exponent in (\ref{eq: 3.3.8}) 
satisfies $\alpha<1$. This implies that the integral defining the mean value is divergent 
in the approximation (\ref{eq: 3.3.8}), and that the mean value (which is finite) 
is determined by the behaviour of $P_S$ far into the tail of its distribution. 
If $0<\alpha<1$ and $k<1$, the value of $\langle \Delta S^k\rangle$ is determined 
by $P_S$ at small values of $S$, so 
that $\langle \Delta S^k\rangle/\langle S\rangle$ is small, without the necessity for a 
delta-function component. 

Hence we can conclude that when 
$\ln \Lambda>\ln \Lambda_{\rm c}=\sigma^2/2$, we always have a $k>1$ such that 
$\xi_k<0$, implying that $P_S$ condenses onto $\delta(S-S_0)$ as $N\to \infty$. 
When $\Lambda<\Lambda_{\rm c}$, we can have $\xi_k<0$, when $k<1$. This however
is just a consequence of the very long tail of the distribution, and it does not imply 
condensation onto a delta function.

\subsection{Sum is dominated by its largest term}
\label{sec: 4.2}

It is possible that the tail of the distribution of $S$ is, in fact, dominated by the largest 
value of $X$, so that when $S\gg S_0$, $P_S$ approaches $P_{X_{\rm m}}$. 
In the following we provide evidence that this is indeed the case. 
First, we note that if $\xi_k>1$, the divergence of $M_k$ as 
$N\to \infty$ will be determined by the tails of the distribution of $S$. 
By making the change of variables $X = \exp(NY)$, and so $X_{\rm m} = \exp(NY_{\rm m})$, 
in the tail of the distribution we have $P_{Y_{\rm m}}\sim {\cal N}\exp[-NJ(Y_{\rm m})]$, so that 
\begin{equation}
\label{eq: 4.2.1}
\langle \Delta S^k\rangle \sim \int {\rm d}Y\ \exp\left[N\left(\ln \Lambda+kY-J(Y)\right)\right]
\ .
\end{equation}
Using the Laplace principle and writing $F(Y)=\ln \Lambda +kY-J(Y)$ we estimate
\begin{equation}
\label{eq: 4.2.2}
\langle \Delta S^k \rangle \sim \exp[NF(Y^\ast)]
\end{equation}
where $F'(Y^\ast)=0$. For the log-normal model we have $Y^\ast=\mu+k\sigma^2$, so that 
\begin{equation}
\label{eq: 4.2.3}
\langle \Delta S^k\rangle \sim \exp\left[N\left(\ln \Lambda +k\mu +k^2\sigma^2/2\right)\right]
\ .
\end{equation}
Combining this estimate with Eqs.~(\ref{eq: 3.1.1}), (\ref{eq: 3.1.2}) and (\ref{eq: 3.2.7}) 
we recover equation (\ref{eq: 3.2.9}), hence suggesting that 
the tails of the distribution of $P_S$ are asymptotic to the 
distribution of the largest element of the sum. Numerical results presented in the next section 
show that this is indeed the case, see Fig.~\ref{fig: 3}. 

\subsection{Nature of the phase transition}
\label{sec: 4.3}

The arguments presented so far imply the existence  of a phase transition, which 
occurs at a critical value $\Lambda_c$. For the log-normal model we have shown this is 
$\Lambda_{\rm c}=\exp(\sigma^2/2)$. To quantify this phase transition we will analyse the 
asymptotic behaviour of the PDF of $S$ as $N\to \infty$.
Consider the predicted form of $P_S$ for the supercritical case, $\ln \Lambda>\sigma^2/2$. 
As $N\to \infty$, the distribution approaches a delta function, but we have 
also seen that the tail is in agreement with the distribution of 
the maximum value $X_{\rm m}$. By making use of 
$P_{Y_{\rm m}}\sim {\cal N}\exp[-NJ(Y_{\rm m})]$ and changing back to $X_{\rm m}$, we write
\begin{eqnarray}
\label{eq: 4.3.1}
P_S& \sim &\delta(S-S_0)+{\cal N}\exp\left[-NJ(Y_{\rm m})\right]
\nonumber \\
& \sim & \delta(S-S_0)+\frac{1}{N}\exp(-ND)\left(\frac{S}{S_0}\right)^{-(1+\alpha)}
\end{eqnarray}
where $D=\alpha Z_0+J(Z_0)-\ln \Lambda$, which is a positive quantity. On the other hand, 
for the subcritical case, $\ln \Lambda<\sigma^2/2$, we predict that 
\begin{equation}
\label{eq: 4.3.3}
P_S\sim \frac{1}{N}\exp(-ND)\left(\frac{S}{S_0}\right)^{-(1+\alpha)}
\ .
\end{equation}

\section{Generalisation}
\label{sec: 5}

Thus far we have derived results using explicit formulae 
for the log-normal model. It is desirable to understand how to 
address the same issues for a general probability distribution of 
the positive factors $x_k$, which allows us to introduce
the auxiliary variable $y_k$, defined by $x_k = \exp(y_k)$.
It is convenient to express the results in terms 
of the cumulant generating function $\lambda(k)$ for the distribution of $Y$, defined 
by 
\begin{equation}
\label{eq: 5.1}
\langle \exp(NkY)\rangle=\exp[N\lambda(k)]
\ .
\end{equation}
Because 
\begin{equation}
\label{eq: 5.2}
\langle \exp(NkY)\rangle=\bigg\langle 
\exp\left(k\sum_{i=1}^N y_i\right)\bigg\rangle=\langle \exp(ky)\rangle^N 
\end{equation}
and $x=\exp(y)$ it follows that 
\begin{equation}
\label{eq: 5.3}
\lambda(k)=\ln\,\langle x^k \rangle
\ .
\end{equation}
The cumulant generating function is a Legendre transform of the entropy 
function for the distribution of $Y$:
\begin{equation}
\label{eq: 5.4}
J(Y)=kY-\lambda(k)
\ ,\ \ \ 
k=J'(Y)
\ .
\end{equation}
We can assume that $J(Y)$ is a convex function, so that, for any value of $Y_1$,  
\begin{equation}
\label{eq: 5.5}
J(Y)\ge J(Y_1)+J'(Y_1)(Y-Y_1)
\end{equation}
and a similar result holds for $\lambda(k)$. Now consider how the results of sections 
\ref{sec: 3} and \ref{sec: 4} generalise.

\subsection{Central moments}
\label{sec: 5.1}

Using (\ref{eq: 3.2.8}) and assuming that the maximum growth exponent 
occurs for $l=0$, we find that the exponents for the 
central moment are given by:
\begin{equation}
\label{eq: 5.1.1}
\ln \xi_k=\lambda(k)-k\lambda(1)-(k-1)\ln \Lambda
\ .
\end{equation}
When $k<2$, only the $l=0$ case need be considered, so that (\ref{eq: 5.1.1})
is certainly valid in that case.
The critical point for the phase transition is where $\xi_{1+\epsilon}=1$
as $\epsilon>0$ approaches zero, that is 
\begin{equation}
\label{eq: 5.1.2}
\lambda'(1)-\lambda(1)=\ln\Lambda_{\rm c}
\ .
\end{equation}
In the non-uniform phase, we can use (\ref{eq: 5.1.2}) together 
with the convexity of $\lambda(k)$ and the positivity of $\ln \Lambda$ to 
establish that (\ref{eq: 5.1.1}) is valid for all $k$.

\subsection{Distribution of maximum element of sum}
\label{sec: 5.2}

The maximum value of $X$ has a power-law distribution 
$P(X_{\rm m})\sim X_{\rm m}^{-(1+\alpha)}$, with the exponent 
given by $\alpha=J'(Z_0)$. Using (\ref{eq: 5.4}), we obtain an implicit 
equation for $\alpha$: we have $\lambda(\alpha)=\alpha Z_0-J(Z_0)$ 
with $\lambda'(\alpha)=Z_0$. Noting that $Z_0=\ln\Lambda+\lambda(1)$, we arrive at 
\begin{equation} 
\label{eq: 5.2.1}
\lambda'(\alpha)-\lambda(1)=\ln\Lambda
\ ,
\end{equation}
which is an implicit equation for $\alpha$. In the case of the log-normal model we 
found that the critical point, i.e., where the delta-function component for the large $N$ 
limit of $P_S$ appears, corresponds to the point at which $\alpha=1$. Equation 
(\ref{eq: 5.2.1}) implies that, in the general case, the condition $\alpha=1$ is satisfied 
at a value of $\Lambda$ which satisfies equation (\ref{eq: 5.1.2}). We conclude that, 
in our model, the delta function distribution occurs whenever the decay of the 
distribution of the largest element is sufficiently rapid that the mean value 
of $S$ is close to the mode of the distribution of $S$.

\subsection{A consistency check}
\label{sec: 5.3}

As a consistency check, we should verify that $Z_0\ge Y_0$, that is, the 
peak of the distribution of the maximum value of $X$ lies below the mean value of $S$,
consistent with equation (\ref{eq: 4.3.1}). 
This ensures that the PDF of the tail of $P_S$ is already exponentially small 
for $S$ 
just slightly greater than $\langle S\rangle$. 

This is true for the log-normal model, where, 
setting $A^2=\ln \Lambda$ and $B^2=\sigma^2/2$, we write $Z_0=\mu +A^2+B^2$ 
and $Y_0=\mu+2AB$. Because $A^2+B^2-2AB=(A-B)^2\ge 0$, we do confirm that 
$Z_0\ge Y_0$, as expected.

It is less easy to see why this should be true in the case of a general distribution of $S$. 
Recalling that $\langle x^k\rangle=\exp[\lambda(k)]$, the values of $Y_0$ and $Z_0$ 
are defined via the following relations
\begin{equation}
\label{eq: 5.3.1}
J(Y_0)=\ln \Lambda
\ , \ \ \ 
Z_0=\ln \Lambda+\lambda(1)
\ .
\end{equation}
Define $Y_1$ to be the image point of $k=1$ under the Legendre 
transformation:
\begin{equation}
\label{eq: 5.3.2}
J(Y_1)=Y_1-\lambda(1)
\ ,\ \ \ 
J'(Y_1)=1
\ .
\end{equation}
Equations (\ref{eq: 5.3.1}) give $Z_0=J(Y_0)+\lambda(1)$, and hence
\begin{equation}
\label{eq: 5.3.3}
Z_0-Y_0=J(Y_0)+\lambda(1)-Y_0=J(Y_0)-J(Y_1)+Y_1-Y_0
\ .
\end{equation}
Noting that $J'(Y_1)=1$, the convexity relation (\ref{eq: 5.4}) then establishes 
that $Z_0-Y_0\ge 0$.

\section{Numerical investigations}
\label{sec: 6}

We investigated the distribution of $S/\langle S\rangle$ for our model
to verify that the phase transition exists as $N\to \infty$, and that it is correctly 
described by our theory. We used the log-normal distribution and the uniform distribution. 

\subsection{Log-normal distribution}
\label{sec: 6.1}

\begin{figure}[t]
\centering
\includegraphics[width=0.99\textwidth]{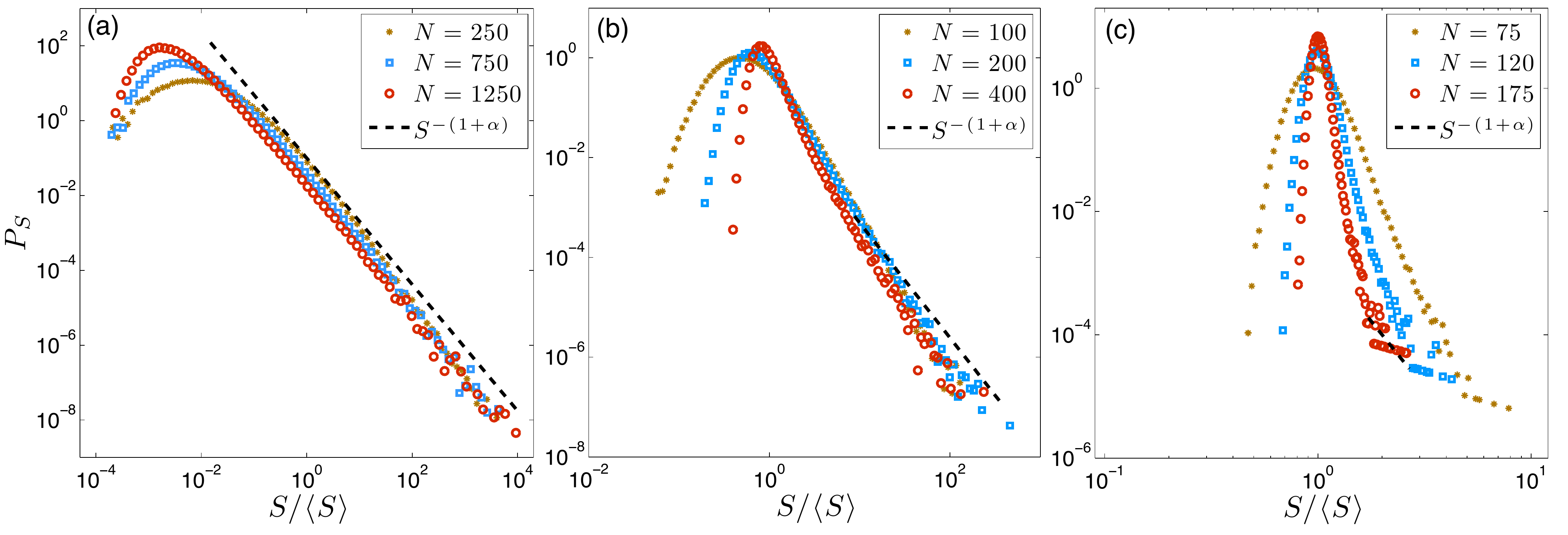}
\caption{PDF of the random variable $S$  when the variables $x_k$ follow a log-normal distribution 
with $\mu=0$ and $\sigma=0.18$, and for $\Lambda=0.99\Lambda_c$ (a), 
$\Lambda=1.01\Lambda_c$ (b), and $\Lambda=1.05\Lambda_c$ (c). 
Different symbols correspond to different values of $N$.
The dashed line is a power-law function that goes as $S^{-(1+\alpha)}$, where 
$\alpha$ is given by Eq.~(\ref{eq: 3.3.7}). } 
\label{fig: 2}
\end{figure}

The explicit calculations for this model have been derived in the previous section, 
where we have computed that the critical value of the phase transition occurs at 
$\Lambda_{\rm c}= \sigma^2/2$. Figure \ref{fig: 2} shows numerical results 
for $\sigma=0.18$ and for different values of $\Lambda$ that are below and above 
the critical point. In the sub-critical case the distribution is approximated by a power-law, 
with an exponent which is independent of $N$, whereas in the super-critical case the distribution 
sharpens as $N$ increases.

\subsection{Uniform distribution}
\label{sec: 6.2}

We also consider the case where the random variable $x_k$ follows  a uniform distribution
in the interval $[ 0,\ell]$. We first use the results of section \ref{sec: 5} to determine 
the critical point $\Lambda_{\rm c}$, and the exponent $\alpha$.

\begin{figure}[t!]
\centering
\includegraphics[width=0.99\textwidth]{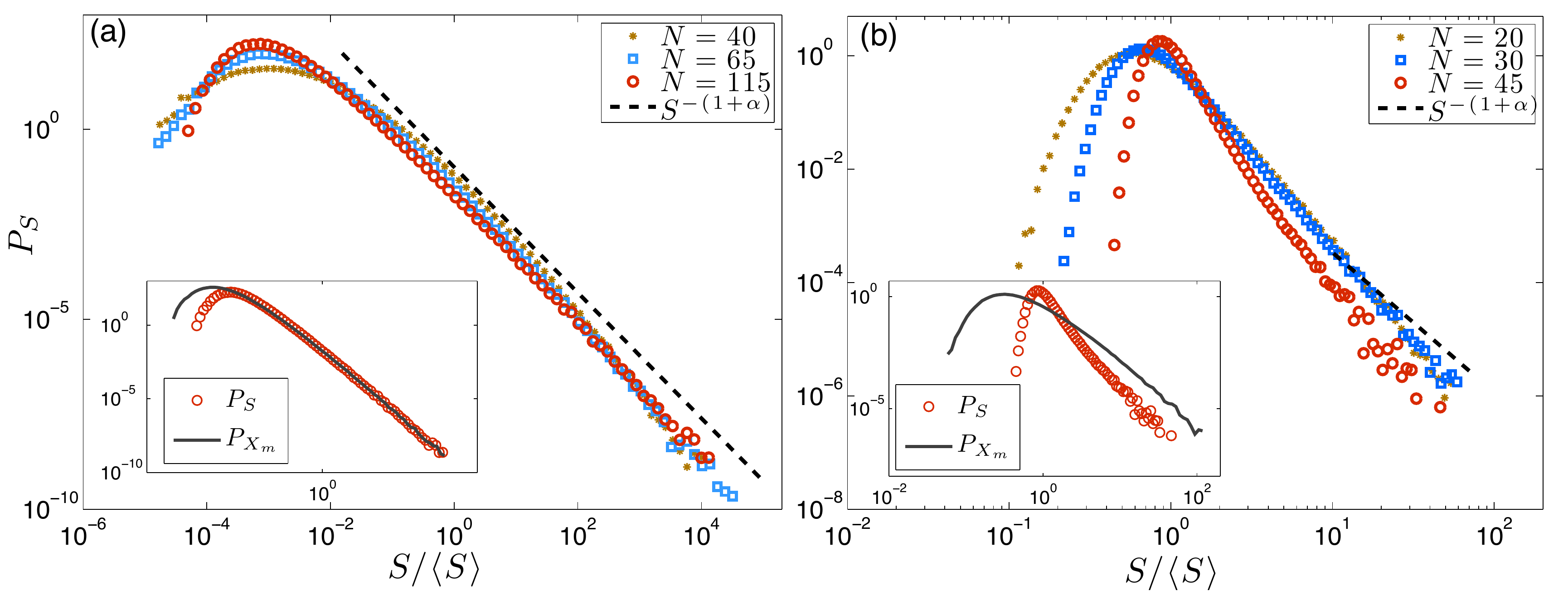}
\caption{PDF of the random variable $S$  when the variables $x_k$ follow  a uniform distribution in $[0,1]$, and for $\Lambda=0.95\Lambda_c$ (a),  and $\Lambda=1.05\Lambda_c$ (b). Different symbols correspond to different values of $N$, and  the dashed line is a power-law function that goes as $S^{-(1+\alpha)}$, where 
$\alpha$ is given by Eq.~(\ref{eq: 6.2.1}). The inset panels show a comparison between $P_S$ and the PDF of the largest element in the sum, $P_{X_{\rm m}}$, shown by the full line.
}
\label{fig: 3}
\end{figure}

To determine explicitly the entropy function $J$, we start with finding 
the moments, which is then used to determine the cumulant via equation (\ref{eq: 5.3}):
\begin{equation}
\langle x^k \rangle = \int_0^\ell {\rm d}x\ x^{k} = \frac{\ell^k}{k+1}
\ ,\qquad
\lambda(k) = k\ln\ell- \ln( 1+k )
\ .
\label{eq:k-mom_x}
\end{equation}
To determine the entropy function, we could adapt  Example 2.3 p.6 of Ref.~\cite{Tou09}, 
or else use (\ref{eq: 5.4}) to express $J(Y)$ as the Legendre 
transform of $\lambda(k)$: $J(Y)=kY-\lambda(k)$, with $\lambda'(k)=Y$. 
(Note that the way things are defined, $Y<\ln\ell$ and $ k > -1$). 
We find $Y=\ln\ell-1/(1 + k)$, and eliminating $k$ 
in (\ref{eq: 5.4}) immediately gives:
\begin{equation}
J(Y) = - Y - 1 +\ln\ell- \ln(\ln\ell-Y)
\label{eq:fn_J}
\end{equation}
which is clearly convex, positive and has a minimum when $Y=\log\ell-1$. 
Using now equation (\ref{eq: 5.2.1}) we determine the exponent for the decay of the 
distribution $P_S\sim S^{-(1+\alpha)}$. We find $\ln \Lambda=\ln(2)-1/(1+\alpha)$ that 
 gives 
\begin{equation} 
\label{eq: 6.2.1}
\alpha=\frac{1}{\ln (2/\Lambda)}-1
\ ,
\end{equation}
which is independent of $\ell$. Setting $\alpha=1$, or 
equivalently applying equation (\ref{eq: 5.1.1}), we find that the critical value is given by 
\begin{equation}
\label{eq: 6.2.2}
\Lambda_{\rm c}=\frac{2}{\exp(1/2)}
\ .
\end{equation}
Figure \ref{fig: 3} shows numerical results for $P_S$  and $P_{X_{\rm m}}$ for 
the case where the $x_k$ have a uniform distribution, with $\ell=1$.

\section{Conclusions}
\label{sec: 7}

We model intensity fluctuations by a sum of an exponentially 
increasing number of path contributions ${\cal N}\sim \Lambda^N$, each of 
which have a multiplicative distribution, with $N$ random factors. Our 
calculations indicate that there is a phase transition, with a critical value 
of $\Lambda$:

\begin{enumerate}

\item The distribution $P_S\sim \delta(S-S_0)$ as $N\to \infty$ when 
$\Lambda>\Lambda_{\rm c}$, apart from a power-law tail, with a coefficient
which becomes exponentially small in the large $N$ limit.

\item When $\Lambda<\Lambda_{\rm c}$, $P_S\sim S^{-(1+\alpha)}$ is approximately
a power-law. with $\alpha<1$.

\item In both cases, there is a tail of $P(S)$ which is asymptotic to the PDF of the largest element
of the sum. This PDF can be obtained analytically.

\end{enumerate}
 
Numerical investigations on two solvable models verify these results, showing 
that there is a transition between a phase where $S$ has a delta-function distribution 
in the limit as $N\to \infty$, and a phase dominated by fluctuations, 
where $S$ has a very broad distribution approximated by a power-law.

Finally, it would be interesting to explore whether the ideas developed in this 
work could shed light on the transition observed in models of hopping 
conductivity~\cite{Ngu+85,Spi+96}. The model studied here could also conceivably
shed light on the phenomenon of concentration of density in models of particles
transport by a compressible flow discussed in \cite{Pradas+17}.

\end{document}